\documentclass{jaa}
\usepackage{natbib}
\def\aap {{A\&A\,}}
\def\aas {{AAS\,}}

\def\actaa {{AcA\,}}
\def\apj {{ApJ\,}}
\def\apjs {{ApJS\,}}
\def\apjl {{ApJL\,}}
\def\aj {{AJ\,}}

\def\araa  {{ARA\&A\,}}

\def\pasp {{PASP\,}}
\def\pasj {{PASJ\,}}

\def\apss {{Ap\&SS\,}}

\usepackage{longtable}
\usepackage{graphicx}
\usepackage{color, colortbl}
\usepackage[colorlinks=true,linkcolor=blue, citecolor=blue]{hyperref}% 

\begin{document}\sloppy

%%paper title
%%For line breaks \\ can be used within title
\title{Identifying the population of T-Tauri stars in Taurus: UV-optical synergy}

%the Gaia, GALEX, and TANSPEC synergy
%%author names are separated by comma (,)
%%use \and before the last author name
%%use a * along with the number separated by comma
%% for the  author for correspondence
%%\textsuperscript{number} is used for affiliation
%%\affilOne, \affilTwo etc., upto \affilTwentyfive is possible
%%Please note the first letter after \affil is capitalised in the command
%%

\author{Prasanta K. Nayak\textsuperscript{1,5*}, Mayank Narang\textsuperscript{1,4}, Manoj Puravankara\textsuperscript{1}, Himanshu Tyagi\textsuperscript{1},Bihan  Banerjee\textsuperscript{1}, Saurabh  Sharma\textsuperscript{2} ,  Rakesh Pandey\textsuperscript{2}, Arun Surya\textsuperscript{1},  Blesson Mathew\textsuperscript{3}, R. Arun\textsuperscript{6}, Ujjwal, K\textsuperscript{3} and Sreeja S Kartha\textsuperscript{3} }
\affilOne{\textsuperscript{1}Department of Astronomy and Astrophysics, Tata Institute of Fundamental Research, Mumbai, 400005, India.\\}
\affilTwo{\textsuperscript{2} Aryabhatta Research Institute of Observational Sciences (ARIES), Manora Peak, Nainital 263 001, India \\}
\affilThree{\textsuperscript{3} Department of Physics and Electronics, CHRIST (Deemed to be University), Bangalore 560029, India \\}
\affilFour{\textsuperscript{4} Academia Sinica Institute of Astronomy \& Astrophysics, 11F of Astro-Math Bldg., No.1, Sec. 4, Roosevelt Rd., Taipei 10617, Taiwan, R.O.C. \\}
\affilFive{\textsuperscript{5} Institute of Astrophysics, Pontificia Universidad Católica de Chile, Av. Vicuña MacKenna 4860, 7820436, Santiago, Chile \\}
\affilSix{\textsuperscript{6} Indian Institute of Astrophysics, 2$^{nd}$ block Koramangala, Bangalore, 560034, India}
% \affilTwo{\textsuperscript{2}Department of Q, University Z, Place Pincode, Country.}

%%escape two column mode for title, affiliation and abstract
%%by giving \twocolumn command as shown

\twocolumn[{

\maketitle

%%include \corres to print the corresponding author Email id
\corres{nayakphy@gmail.com}

%%include \msinfo for
%%manuscript information such as
%%received, revised and accepted dates
%%
%\msinfo{1 November 2022}{25 July 2023}
%MS received 1 November 2022; accepted 25 July 2023

%%abstract
\begin{abstract}
With the third data release of the Gaia mission  $Gaia$ DR3 with its precise photometry and astrometry, it is now possible to study the behaviour of stars at a scale never seen before. In this paper, we developed new criteria to identify T-Tauri stars (TTS) candidates using UV and optical CMDs by combining the GALEX and Gaia surveys. We found 19 TTS candidates and 5 of them are newly identified TTS in the Taurus Molecular Cloud (TMC), not catalogued before as TMC members. For some of the TTS candidates, we also obtained optical spectra from several Indian telescopes. 
We also present the analysis of the distance and proper motion of young stars in the Taurus using data from Gaia DR3.  We found that the stars in Taurus show a bimodal distribution with distance, having peaks at $130.17_{-1.24}^{1.31}$ pc and $156.25_{-5.00}^{1.86}$ pc. The reason for this bimodality, we think, is due to the fact that different clouds in the TMC region are at different distances.  We further show that the two populations have similar ages and proper motion distribution.  Using the $Gaia$ DR3 colour-magnitude diagram, we show that the age of Taurus is consistent with 1 Myr. 
\end{abstract}

%%insert keywords separated by 3 hyphens using \keywords{words}

}]
%%close the twocolumn escape here

%%include \doinum{number}for the DOI number in the header
%%include \volnum{number} for the volume number in the header
%%include \year{yyyy} for  year of publication in the header
%%include \pgrange{num--num} page range of article in the header
%%include \artcitid{num} for the article citation id
%%include \lp to print last page of the article
%%include \setcounter{page}{pagenum} for the exact starting page of the article

\doinum{12.3456/s78910-011-012-3}
\artcitid{\#\#\#\#}
\volnum{000}
\year{0000}
\pgrange{1--}
\setcounter{page}{1}
\lp{1}

% Section 2: 

% Sec 3: 

\section{Introduction}
Taurus-Auriga molecular cloud complex (TMC in short) is one of the nearest dark star-forming regions where low-mass stars are being formed. The TMC spans about  15$\times$15 square degrees on the sky \citep{Esplin14, Esplin17}. TMC is an ideal region to study star formation from the deeply embedded source, i.e., proto-stars that are at the early stages of evolution to disk-free Class III source at the end of their pre-main-sequence lifetime and are thought to be the precursor to planetary systems \citep{Kenyon08, Furlan07, Furlan11}. TMC does not have O or B-type stars and contains about 10$^4$ M$_\odot$ of molecular gas. The cloud has a filamentary structure which was identified through surveys in $^{12}$C$^{16}$O, $^{13}$C$^{16}$O and OH surveys \citep{Kenyon08}.

There are several studies that have been carried out to estimate the membership of the TMC and to understand the pre-main-sequence populations of the region \citep[e.g.,][]{Elias78, Bertout99, Kenyon08, Furlan07, Furlan11, Rebull11, Luhman11, Luhman2018, Galli18, Galli19}. 
However, there are only a few dedicated studies to search and characterize the T-Tauri Stars (TTS).  TTS are low-mass pre-main-sequence (PMS) stars, generally categorized into Classical TTS (CTTS) and Weak-line TTS (WTTS) based on their H$\alpha$ equivalent widths \citep{Bertout1989}. CTTS show strong H$\alpha$ emission indicating active ongoing accretion from the circumstellar disk to the central star, while non-accreting WTTS show weak emission lines, suggesting weak or no accretion.  The study of TTS is fundamental for our understanding of the formation and evolution of the solar system. % and planetary build-up.

The accretion process in CTTS causes the release of excess energy, which appears as excess emission in the UV continuum regions of spectral energy distribution (SED) as well as in the spectral indicators (C~II, S~II, Fe~II, Mg~II, C~IV, and the H$_2$ molecular emission \citep{Gomez2009_uv}. The excess UV continuum and UV spectra are the primary indicators of the accretion process in CTTS. 
The CTTSs also emit excess emission in infrared to millimeter wavelengths, caused by radiation from the disk.  Non-accreting WTTS also show UV excess due to their chromospheric activity but are significantly low compared to CTTS. Therefore, by constructing the spectral energy distributions (SEDs) of stars by combining UV photometry with available optical/IR photometric data, one can not only distinguish the CTTS from WTTS but also identify new TTS candidates and characterize them.

Most of the previous studies in search of TTS candidates are based on secondary indicators, such as the equivalent width of the H$\alpha$ emission line or the presence of enhanced flux in the infrared region of SEDs \citep{Furlan11}. Only a few studies have also searched TTS in the UV \citep{FH10, Gomez2015}. 
However, all of these studies used a fixed distance to all the members of the TMC and did not have information on the proper motions of the members.

Most of the studies identified TMC to be located at a distance of 140 pc \citep[e.g.,][]{Elias78, Kenyon08, Furlan07, Furlan11}. However, later studies identified TMC to be more extended, possibly from 125 to 168 pc \citep{Bertout99, Torres2007, Torres2009, Galli18, Fleming2019}. The distance to TMC has been estimated through different methods such as star counts \citep[145 pc;][]{Greenstein1937}\citep[142 pc;][]{McCuskey1939}, photometry of bright stars associated with the reflection nebulae \citep[135$\pm$10 pc;][]{Racine1968}, reddening turn-on method using stars associated with Lynds dark clouds \citep[d = 140--175 pc;][]{Straizys1980, Meistas1981} and parallax measurements comprising spectroscopic parallax \citep[140$\pm$10 pc;][]{Kenyon1994} and trigonometric parallax with $Hipparcos$ \citep[$139_{-9}^{10}$ pc;][]{Bertout99} and VLBI \citep[d$_{near}$ = 126.6$\pm$1.7 pc and d$_{far}$ = 162.7$\pm$0.8 pc;][]{Galli18, Galli19}.

Thanks to the Gaia survey, which provides accurate estimation of parallax and proper motions with less than a few per cent in error. With the help of Gaia, the number of TMC members is now increased to over 500 from a couple of hundred \citep{Luhman2018, Fleming2019, Esplin2019}. 
Recently, using Gaia´s second data release (Gaia DR2) \citep{GaiaCollaboration2018}, it was reported that the sources in TMC show a bimodal distribution in the distance \citep{Fleming2019}. The two populations peak at 130.6$\pm$0.7 pc and 160.2$\pm$0.9 pc, respectively.  \cite{Fleming2019} further showed that these two populations have different kinematic behavior as well, with a mean value in proper motion at 24.5$\pm$2.8 and 20.1$\pm$2.4 mas yr$^{-1}$, respectively.

{ Gaia$'$s data release 3 (Gaia DR3) brings a significant advancement compared to Gaia DR2 in terms of precision improvement in photometry, systematic errors, and astrometric solutions. Precision in parallax is increased by 30 percent, in proper motions it is improved by a factor of 2. The systematic errors in the astrometry were suppressed by 30-40\% for parallax and by a factor of $\sim$2.5 for proper motions. Gaia DR3 also provides better precision in photometry which is homogeneous across color, magnitude, and celestial position with no systematics above the 1\% level in any passbands (G, BP, RP) \citep{GaiaCollaboration2021}.
As discussed earlier, the presence of UV excess in a PMS star indicates that the star is a TTS, where UV excess is generated from the accretion process or chromospheric activity. Therefore, UV observation of PMS stars and comparing them with their optical observation will help us to identify the TTS, and measuring the UV excess tells us about their accretion or chromospheric properties. GALEX GR6/7 is the most recent catalog of all-sky UV surveys, provides photometry as deep as 22 mags in AB systems in far-UV (FUV, $\lambda_{eff}$ $\sim$ 1528 \AA) and near-UV (NUV, $\lambda_{eff} \sim$ 2310 \AA) passbands \citep{Bianchi2017} with a spatial resolution of 5$''$. There have been previous attempts to use an older version of GALEX GR5 data \citep{Bianchi2014} to identify TTS by \cite{Gomez2015}, where authors used UV-IR color-color diagrams to separate TTS from other stars. However, due to the absence of the Gaia survey, authors ended up finding many TTS candidates that are not part of Taurus. At the same time, the authors categorized some CTTS as WTTS (for example V836 Tau) based on the color-color diagram, which shows a strong signature of FUV line emissions (like C IV) \citep{Ingleby2013}.   
}

In this work, we have searched for new TTS candidates in the TMC by combining Gaia-DR3 \citep{GaiaCollaboration2022} and GALEX GR6/7 \citep{Bianchi2017}. { We also carry out follow-up optical-IR spectroscopic observations of some of the newly identified TTS using India's ground-based telescopes and measure the strength of H$\alpha$ emission to confirm their candidature. As mentioned earlier, accretion is the main reason for the excess UV emission in the accreting TTS, and non-accreting TTS emit relatively less UV emission. Therefore, if we can measure the UV luminosity in TTS, we can characterize the TTS whether they are accreting or non-accreting based on their excess UV emission over photospheric emission. On the other hand, the emitted UV luminosity in accreting TTS will give a direct measure of accretion luminosity. Therefore, using follow-up spectroscopic observation, we will be able to establish a relation between UV luminosity (a primary indicator of accretion luminosity) and H$\alpha$ luminosity (a secondary indicator of accretion luminosity). 

As discussed above, the distance to the TMC is not yet well established. Different studies have reported different values of distance. A recent study by \citep{Fleming2019} using Gaia-DR2 reported that the distribution of TMC members has two peaks with two different kinematics. Now, this finding gives birth to the following questions whether TMC is a combination of two different clouds, or both are part of the same molecular cloud but falling apart due to tidal interaction, and whether both the populations have the same star formation history. With the improvement in the astrometric solution in Gaia-DR3, we have also re-analyzed the bimodality of TMC using a more robust and larger (by a factor of 2) sample of stars as compared to \citep{Fleming2019} and try to address the above questions.} In Sect. 2, we describe our sample and the process of retrieving the $Gaia$ DR3 and GALEX data. In Sect. 2.1, we discuss the following spectroscopic observations of newly identified TTS candidates. In Sect. 3, we present the results on the distance and proper motion distribution of stars in TMC. We also discuss various binning techniques and the caveats in interpreting the histograms when data is binned at various intervals.  This puts the results from our analysis and a very strong statistical footing as compared to previous distance estimates. The major results of this study are summarized in Sect. 4.

\section{Identification of new Taurus candidates} \label{new_TTS_data}

In search of new TTS candidates, we first select $\sim$15$\times$15 square degree GALEX GR6/7 \citep{Bianchi2017} data around the TMC regions. Then we restrict the data where both FUV and NUV observation are there. We also remove sources if there are any FUV or NUV artifacts in the data. Then we cross-match the GALEX data with Gaia-DR3 and look for the closest match with a search radius of 3$''$.%, which is the angular resolution of GALEX. 
We put constraints on some of the Gaia parameters and removed the sources with higher $astrometric \, excess \, noise$ ($>$1.3) and higher $RUWE$ ($>$10). We are left with 8846 sources.
We also cross-match the data with the Gaia-DR3-distance catalog by \citet{Bailer2021} to get the distance to each source. 
We consider that the extent of the TMC is between 100 to 200 pc and remove all the sources having distance outside this distance range from the cross-matched catalog. Finally, we ended up with 330 sources within 100 to 200 pc.
 %In the section \ref{method}8846.

%%%================================

\begin{figure*}[h]
\centering
\includegraphics[width=0.45\linewidth]{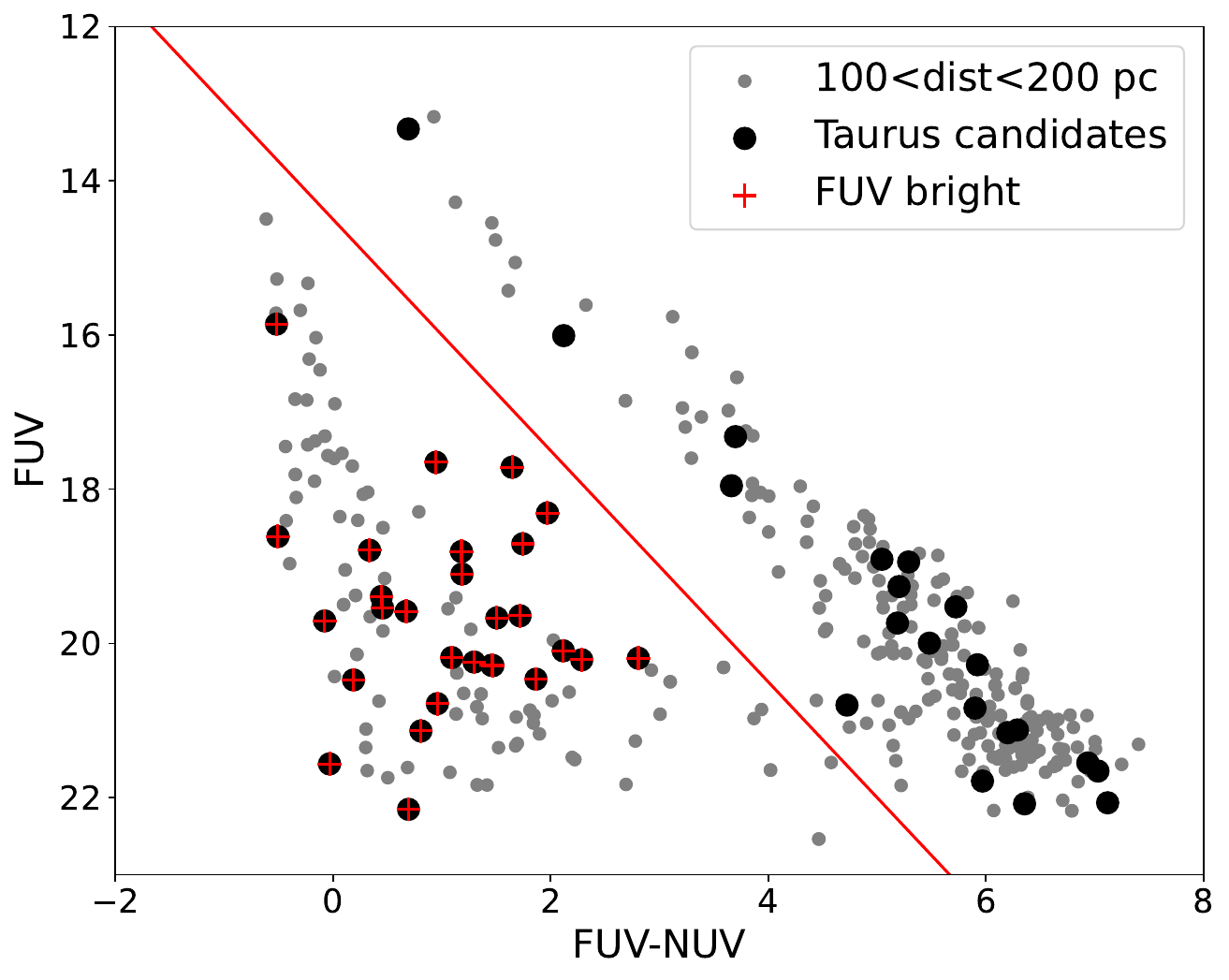}
\includegraphics[width=0.45\linewidth]{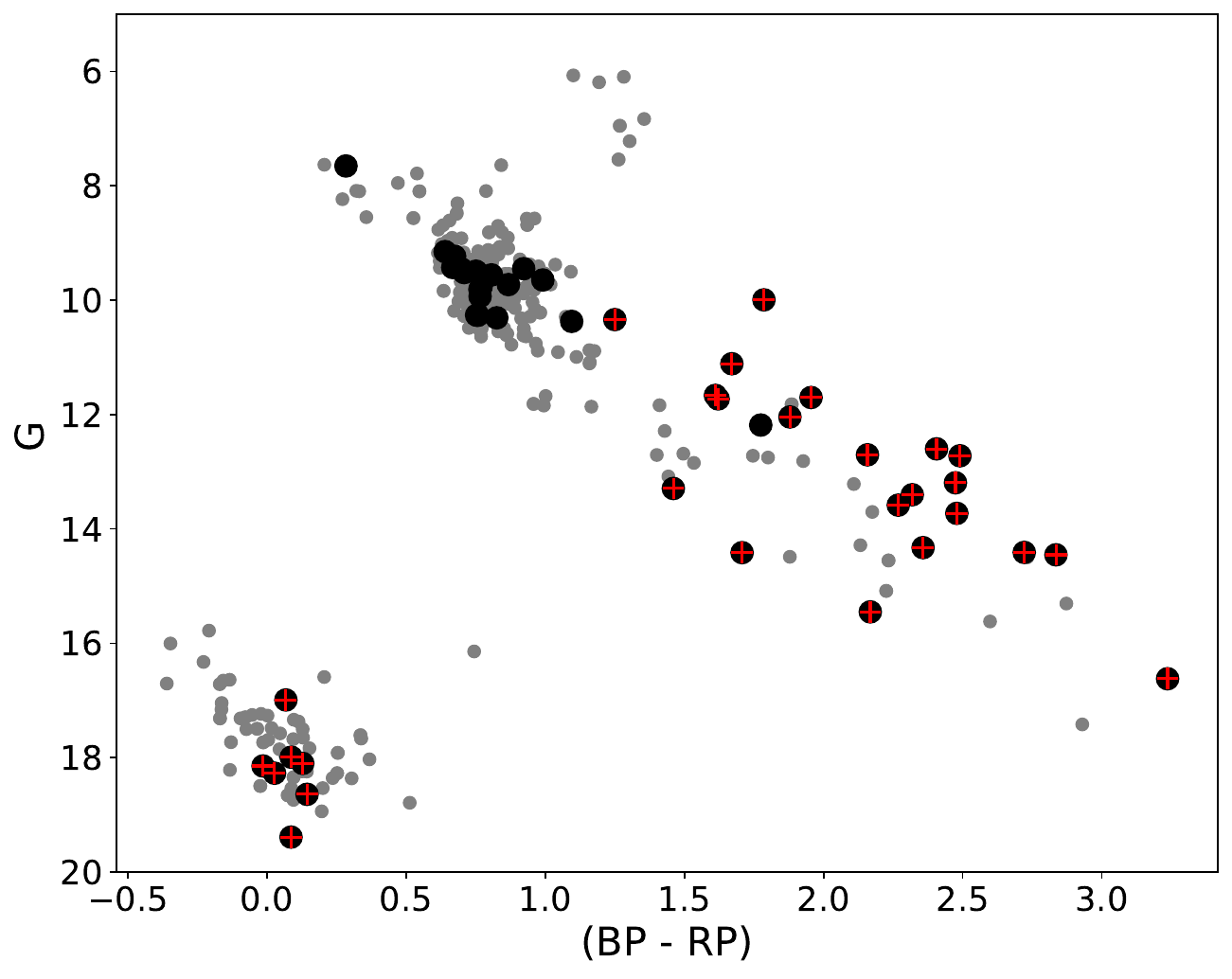}
\caption{UV (left) and optical CMD (right) of sources with GALEX observations along the TMC. The sources with 100 to 200 pc are marked as grey, while the sources with a proper motion similar to that of TMC members are marked as black points. The red line in the UV CMD separates hotter and FUV bright TMC members, marked as a red plus. The same FUV bright populations are located in the WD and PMS regions of the optical Gaia CMD. The FUV bright sources located in the PMS region are nothing but the TTS with excess flux in the FUV due to the accretion process. }
\label{new_tts}
\end{figure*}

%%%% NEW TTS candidates
\begin{table*}
\tabcolsep2.95pt 
\centering
\caption{The table represents the list of 19 TTS identified using our method by combining GALEX GR6/7 and Gaia-DR3}
\begin{tabular}[]{|c|c|c|c|c|c|c|c|c|c|c|c|}
\hline
% \multicolumn{7}{|c|}{}\\
% \multicolumn{7}{|c|}{Sample from \cite{Esplin19}}\\
% \multicolumn{7}{|c|}{}\\
% \hline
RA & Dec & FUV & FUV$_{err}$ & NUV & NUV$_{err}$ & Plx & Plx$_{err}$ & pmra & pmra$_{err}$ & pmdec & pmdec$_{err}$ \\
(deg)  & (deg) & mags & mags & mags & mags  & mas & mas & mas/yr & mas/yr & mas/yr & mas/yr \\\hline
61.714059 &	25.691264 & 20.1 & 0.15 & 17.98 & 0.04 &  6.296 & 0.036 & 12.204 & 0.044 & -17.973 & 0.026  \\
62.418896 &	18.752410 & 21.56 & 0.42 & 21.6 & 0.37 &  8.285 & 0.085 & 4.178 & 0.097 & -13.201 & 0.073  \\
63.340774 &	18.173326 & 22.15 & 0.42 & 21.46 & 0.25 &  8.090 & 0.028 & 3.544 & 0.031 & -13.581 & 0.025  \\
63.553892 &	28.203310 & 20.21 & 0.22 & 17.93 & 0.03 &  8.326 & 0.131 & 6.54 & 0.163 & -27.792 & 0.11  \\
63.556643 &	28.213549 & 19.64 & 0.15 & 17.92 & 0.03 &  7.575 & 0.024 & 8.383 & 0.03 & -24.54 & 0.02  \\
63.697142 &	26.773903 & 20.78 & 0.22 & 19.82 & 0.1 & 7.844 & 0.026 & 8.99 & 0.033 & -22.658 & 0.021  \\
63.699467 &	26.802960 & 21.14 & 0.26 & 20.33 & 0.14 & 7.890 & 0.020 & 8.745 & 0.026 & -22.53 & 0.017  \\
67.464850 &	26.112366 & 20.46 & 0.19 & 18.6 & 0.05 &  7.604 & 0.036 & 5.951 & 0.045 & -21.309 & 0.036  \\
67.684368 & 26.023442 & 18.71 & 0.08 & 16.96 & 0.02 &  7.575 & 0.050 & 6.315 & 0.056 & -21.434 & 0.047  \\
68.630425 &	17.372167 & 20.18 & 0.2 & 19.09 & 0.07 &  6.863 & 0.035 & 12.123 & 0.046 & -20.087 & 0.03  \\
%*71.716 & 13.3674 & 19.58 & 0.11 & 18.91 & 0.06 & 5.58 & 0.025 & 3.783 & 0.031 & -29.434 & 0.02  \\
72.947481 &	30.786982 & 18.31 & 0.05 & 16.34 & 0.02 &  6.567 & 0.037 & 5.238 & 0.046 & -25.917 & 0.032  \\
73.795776 &	30.366385 & 17.65 & 0.05 & 16.7 & 0.02 &  6.325 & 0.049 & 3.748 & 0.059 & -24.298 & 0.035  \\
73.904063 &	30.298532 & 20.19 & 0.16 & 17.39 & 0.03 &  6.360 & 0.021 & 4.509 & 0.024 & -24.177 & 0.014  \\
73.953558 &	17.129578 & 15.86 & 0.02 & 16.38 & 0.02 & 5.636 & 0.057 & 13.803 & 0.072 & -14.943 & 0.049  \\
74.008471 &	30.350913 & 19.39 & 0.16 & 18.95 & 0.07 &  6.053 & 0.082 & 5.317 & 0.109 & -23.992 & 0.065  \\
74.762711 & 30.049984 & 19.67 & 0.14 & 18.16 & 0.04 &  6.380 & 0.017 & 4.533 & 0.017 & -24.641 & 0.013  \\
75.012883 &	30.018997 & 19.1 & 0.1 & 17.92 & 0.03 & 6.247 & 0.062 & 4.696 & 0.068 & -23.608 & 0.048  \\
75.228523 &	32.487903 & 20.29 & 0.18 & 18.82 & 0.05 &  6.001 & 0.026 & 6.849 & 0.027 & -26.475 & 0.019  \\
76.945728 & 31.338377 & 17.72 & 0.05 & 16.07 & 0.01 &  6.313 & 0.019 & 4.129 & 0.019 & -24.458 & 0.014  \\
%81.0331 & 25.7122 & 20.24 & 0.2 & 18.95 & 0.06 &  6.340 & 0.029 & 3.62 & 0.038 & -25.838 & 0.026  \\
\hline
\end{tabular}
\label{Table_tts}
\end{table*}

The left panel of \autoref{new_tts} shows the UV colour-magnitude diagram (CMD) of 330 sources in grey, while the right panel shows the Gaia CMD of the same sample. 
Comparing the recent catalogue of Taurus populations \citep{Esplin2019} with Gaia DR3, we get the range in proper motion values for the Taurus members, discussed in detail in the \autoref{proper_motion}. 
After applying the above proper motion criteria to the sample, we are finally left with 48 sources, which are members of the TMC having both FUV and NUV observations. We also have over-plotted these sources (marked as black points) in the UV and optical CMDs in \autoref{new_tts}. In the UV CMD, we clearly see two different populations. The sources bluer than the red line represent hotter and FUV bright populations. We marked the FUV bright TMC members as a red plus. If we locate these FUV bright populations in the optical CMD, we notice that a few are distributed in the WD region, while the rest are found in the PMS phase. The FUV bright populations located in the PMS phase are the probable candidates for TTS in the TMC. We find that there are 19 such TTS candidates. We have listed these candidates in \autoref{Table_tts}.

We have compared our TTS catalogue with the catalogue given by \citet{Gomez2015}, where the authors have listed 63 new TTS candidates along with the previously known ones. However, distance and proper motion criteria indicate that most of their newly identified TTS are not a part of TMC, and only 23 of them follow the distance and proper motion criteria. However, we found only 7 sources in common with the previously known TTS catalogue by \citet{Gomez2015}. This is because we removed the sources with higher $astrometric \, excess \, noise$ and higher RUWE. We have also compared with the recent catalogue of Taurus members by \cite{Esplin2019} and found 14 sources common with them; however, only three of them are categorized as TTS. 
Hence, in this study, we are able to identify 19 TTS candidates, and only 8 of them are categorized previously in the literature. We found 11 new TTS candidates in the TMC, out of which 5 are added as Taurus members for the first time.

\subsection{Spectroscopic followup observations of TTS candidates}
\begin{figure*}[h]
\centering
\includegraphics[width=0.46\linewidth]{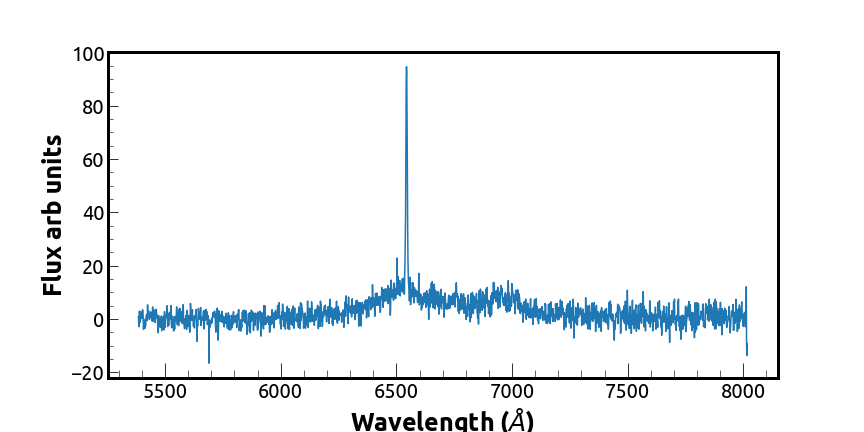}
\includegraphics[width=0.44\linewidth]{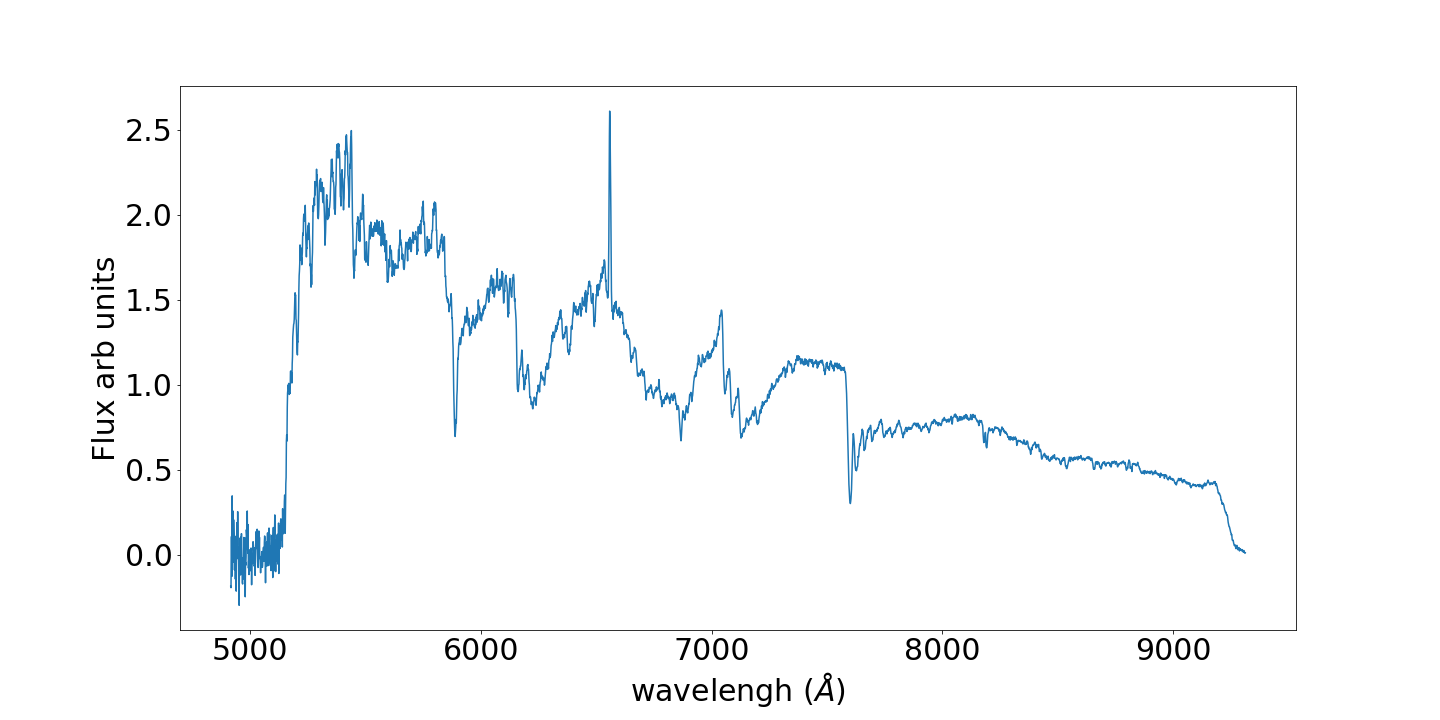}%{figures/HD30171_oct22_SPF_conf.png}
\caption{The Figure shows the optical spectra of two TTS candidates. The left spectrum (RA:68.630425, Dec:17.372167 ) shows strong H$\alpha$ emission indicating the source to be a CTTS obtained using DOT, while the spectra right panel (RA:75.228523, Dec: 32.487903) indicates the source to be a WTTS with weak H$\alpha$ emission obtained from HCT. The right-hand spectrum is one of five new TTS members to Taurus.}
\label{spectra}
\end{figure*}

To further confirm the candidature of these newly identified TTS, identify them as CTTS or WTTS, and study their accretion properties, we have also started obtaining optical-near IR spectra of these sources using the TIFR-ARIES Near-infrared Spectrometer (TANSPEC) in the 3.6m Devasthal Optical Telescope (DOT) and Hanle Faint Object Spectrograph Camera (HFOSC) in the 2m Himalayan Chandra Telescope (HCT). However, we plan to obtain spectra of all the 19 candidates, in order to confirm the candidature of previously TTS. { The TANSPEC has a resolving power of R $\sim$ 2750 and wavelength coverage from 400 to 2500 nm approximately \citep{sharma_2022}. HFOSC works in the wavelength range from 350 to 1050 nm with a range of resolving power (R$\sim$150 to R$\sim$4500) based on the alignment of the grism and slit \footnote{https://www.iiap.res.in/iao/hfosc.html}. For this study, we used grism 8 with a slit configuration of 167l ($sim$ slit width of 1$''$) and resolving power of R$\sim$2000. }
In cycle 2022-C1, we were able to obtain the spectra of two sources using DOT on 29 November 2021. For both sources, we used six frames of observation with six minutes each (3 minutes in dither A and 3 minutes in dither B position). Therefore, the total exposure time was 36 minutes per source. 
We also obtained spectra of four TTS candidates using HCT/HFOSC in cycle 2021-C1. The observational details of DOT and HCT are tabulated in \autoref{Table_new_tts}. 
The optical part of one of the TANSPEC spectra is shown in the left panel of \autoref{spectra}. The spectrum in the left panel shows the presence of strong H$\alpha$ emission, which indicates the source to be a CTTS. One of the HCT spectra is shown in the right panel of \autoref{spectra}, which indicates the source to be a WTTS with weak H$\alpha$ emission. In the upcoming observational cycle, we plan to obtain the spectra of the rest of the TTS candidates using DOT and HCT. Once we have the spectra of all the sources, we will estimate the equivalent width of H$\alpha$ emission to categorize the sources as CTTS or WTTS and estimate their mass accretion rate. We also plan to establish a relation between UV luminosity (a primary indicator of mass accretion) and H$\alpha$ luminosity (a secondary indicator of mass accretion).

%%%% NEW TTS candidates
\begin{table*}
\tabcolsep4.5pt $ $
\centering
\caption{The table represents the details of spectroscopic observation of six TTS candidates out of the 19 sources listed in the \autoref{Table_tts}.}
\begin{tabular}[]{|c|c|c|c|c|c|c|}
\hline
% \multicolumn{7}{|c|}{}\\
% \multicolumn{7}{|c|}{Sample from \cite{Esplin19}}\\
% \multicolumn{7}{|c|}{}\\
% \hline
RA & Dec & observation  & single exposure & number of  & telescope & instrument  \\
(deg)  & (deg) & date & time (mins) & frames &  &  \\\hline
68.630425 &	17.372167 & 29-11-2021 & 6 & 6 & DOT & TANSPEC   \\
%*71.716 & 13.3674 &  29-11-2021 & 36 & DOT & TANSPEC   \\
73.953558 &	17.129578 &  29-11-2021 & 6 & 6 & DOT & TANSPEC  \\
63.553892 &	28.203310 & 17-01-2021 & 10 & 1 & HCT & HFOSC \\
63.556643 &	28.213549 & 17-01-2021 & 20 & 3 & HCT & HFOSC \\
75.012883 &	30.018997 & 18-01-2021 & 45 & 3 & HCT & HFOSC \\
75.228523 &	32.487903 & 17-01-2021  & 20 & 3 & HCT & HFOSC \\
\hline
\end{tabular}
\label{Table_new_tts}
\end{table*}

\section{New look at Taurus population}
\label{sec:data}

\begin{figure}[h]
\centering
\includegraphics[width=0.5\textwidth]{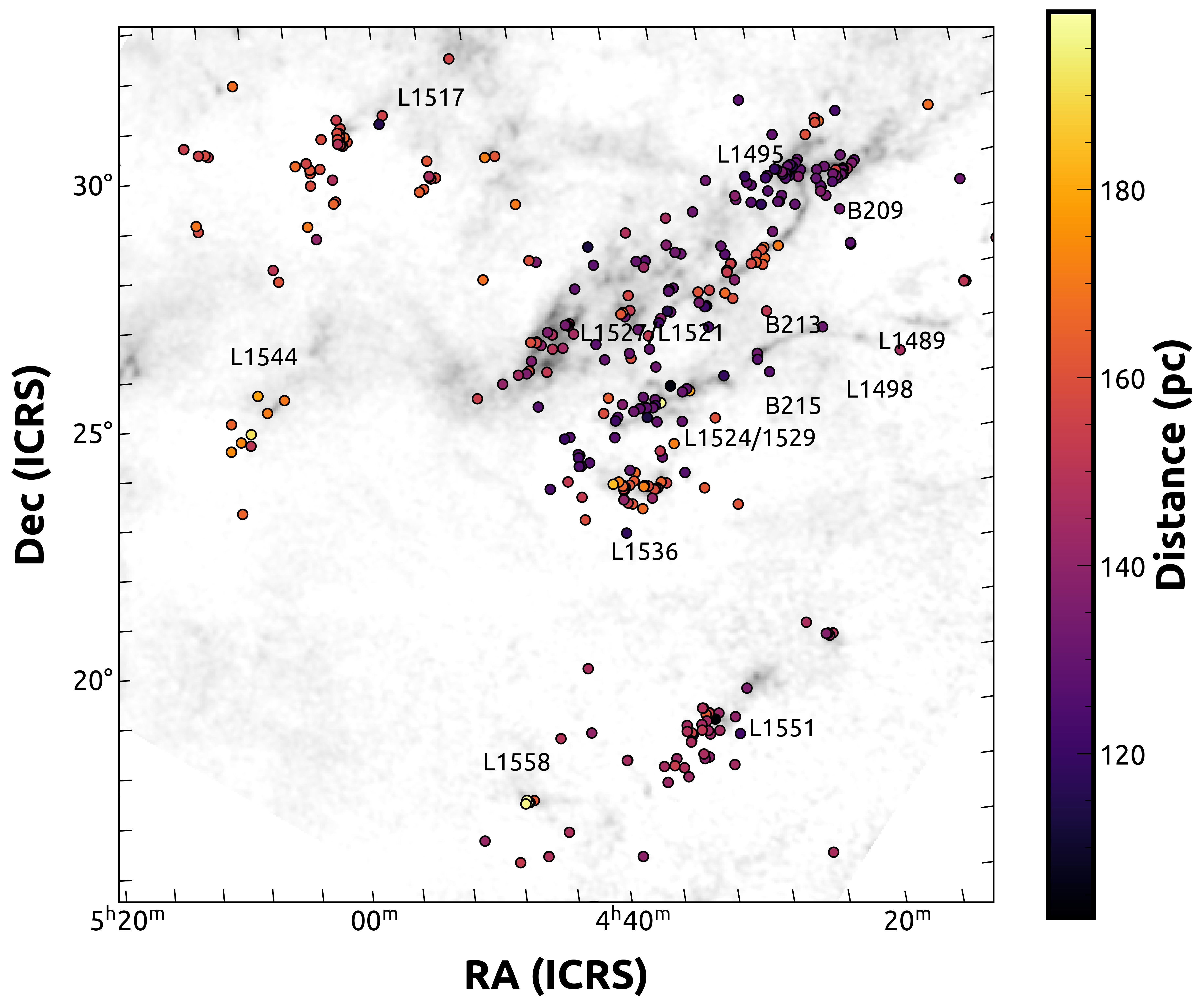}
\caption{The spatial distribution of the \cite{Esplin2019} sample with the distance to each of the stars color-coded and overplotted on the extinction map of TMC from \cite{Dobashi05}. The major Lynd clouds and Bernard clouds are also marked.}
\label{Fig1}
\end{figure}

To ensure that we had the most comprehensive sample of stars in TMC, we used two different catalogs of TMC stars for analysis. The first sample of 519 stars was taken from \cite{Esplin2019}, and the second sample of 215 stars was taken from \cite{Rebull11}. In order to obtain accurate distances to these stars, we cross-matched these samples with $Gaia$ DR3. Out of the 519 stars in \cite{Esplin2019}, we find a Gaia counterpart for 453 stars. Similarly, out of the 215 stars in \cite{Rebull11}, we find a Gaia counterpart for 194 stars. The spatial distribution of the \cite{Esplin2019} sample is shown in \autoref{Fig1} over-plotted on the extinction map of TMC from \cite{Dobashi05}.

In this section, we have analyzed the spatial distribution of young stars in TMC. We have analyzed various optimal binning methods for identifying the features in a distribution in the  \autoref{bin_size}. We show that there is a bimodality in the distribution of stars in TMC in  \autoref{bimodality}.  We further analyzed the proper motion of these stars in \autoref{proper_motion}. This is used to assess whether any morphological or dynamical distinction is seen between the sample of stars in the near and far end of TMC as claimed by \cite{Fleming2019}, discussed in \autoref{classII_III}.  Finally, a $Gaia$ color-magnitude diagram of the sample is presented to ascertain the age of stars in TMC in  \autoref{gaia_cmd}.  %In the following subsections, the details of our analysis are provided.

\subsection{Identification of optimal bin size for a distribution} \label{bin_size}

Optimal binning of data is one of the crucial aspects of data-driven astronomy. When the distribution of data points is represented in the form of a histogram, the number of bins can influence our interpretation of data. By reducing the number of bins considerably, we are, in a way smoothing the distribution and thereby degrading the information in the data. Instead, if we increase the bin number, artificial trends start appearing in the distribution. This can lead to a wrong interpretation of the science case. 

Over the years, various methods have been proposed for estimating the number of bins for a given data sample (also see \cite{Legg13}). Unfortunately, of all these methods, there is no single method that is universally recognized as the best approach for bin size selection. { There are five major methods used to estimate the bin size: Sturges'rule, Rice rule and the rule of square roots, Doane's rule, Scott's rule, and Freedman-Diaconis rule. Details about these methods are mentioned in the Appendix. Each of the binning rules has its own merits and demerits. For our analysis, we employed each of these binning methods on the two different samples of stars in TMC and the results are discussed in the next section.  
}

%%%%====================================

% 

%Figure 1
\begin{figure*}
\centering
\includegraphics[width=1\linewidth]{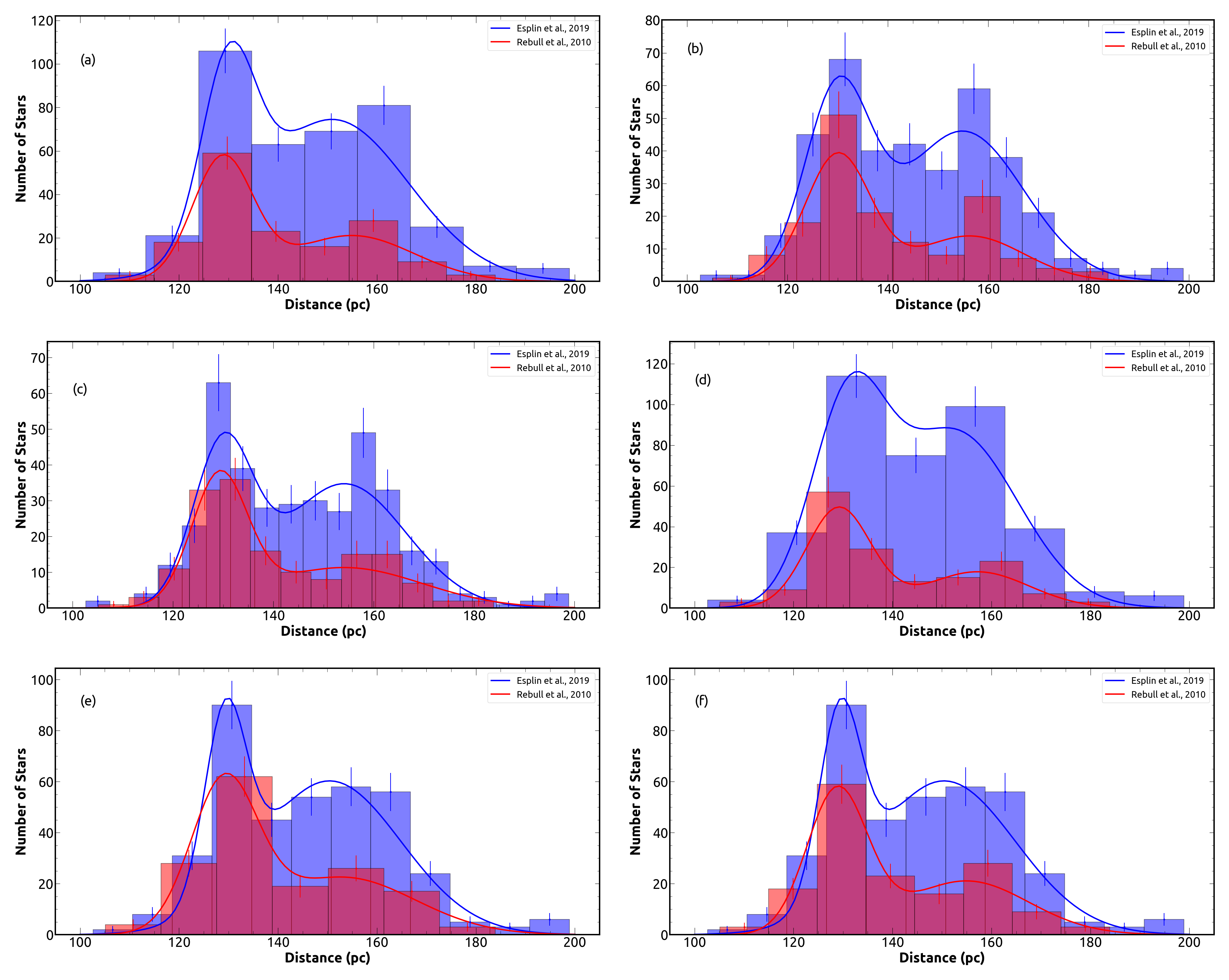}
\caption{ In each of the sub-figures, the distribution of distance for the stars from the two samples, (\cite{Esplin2019} blue, \cite{Rebull11} red) is shown as the histogram using the various binning schemes are shown: (a) Sturges' rule, (b) Rice rule, (c) rule of square root, (d) Doane's rule, (e) Scott's rule, and (f) Freedman-Diaconis rule. Also shown are the fits to the histograms as solid lines.}
\label{Fig2}
\end{figure*}
%%%%%=================================
\begin{table*}
\centering
\caption{The distribution of fit parameters for \autoref{Fig2}}
\begin{tabular}[]{|c|c|c|c|c|c|c|c|}
\hline
\multicolumn{8}{|c|}{}\\
\multicolumn{8}{|c|}{Sample from \cite{Esplin2019}}\\
\multicolumn{8}{|c|}{}\\
\hline
Method & Number of bins & $\mu_1$ & $\sigma_1$ & A1 & $\mu_2$ & $\sigma_2$ & A2 \\
 &   & pc & pc &  & pc & pc & \\\hline
Sturges’ rule & 9 & $129.89_{-2.26}^{13.08}$ & $5.41_{-2.47}^{10.06}$ & $79.96_{-17.10}^{22.46}$ & $151.00_{-3.01}^{7.3}$ & $15.44_{-5.41}^{1.84}$ & $74.46_{-36.23}^{7.95}$ \\
Rice rule & 15 & $129.94_{-1.09}^{1.24}$ & $6.49_{-1.14}^{0.97}$ & $56.92_{-8.99}^{8.19}$ & $154.91_{-3.74}^{2.35}$ & $12.20_{-2.03}^{2.68}$ & $46.03_{-4.37}^{4.55}$ \\
Square root & 20 & $129.89_{-1.01}^{1.08}$ & $5.88_{-1.84}^{1.18}$ & $43.90_{-6.54}^{7.21}$ & $154.12_{-3.79}^{2.53}$ & $12.29_{-1.94}^{2.51}$ & $34.76_{-3.13}^{3.47}$ \\
Doane’s rule & 8 & $130.99_{-2.16}^{11.61}$ & $7.40_{-2.30}^{7.02}$ & $92.56_{-29.42}^{19.23}$ & $152.50_{-4.48}^{4.43}$ & $12.76_{-3.43}^{3.01}$ & $86.79_{-22.65}^{10.84}$ \\
Scott’s rule & 12 & $129.33_{-1.24}^{1.31}$ & $4.20_{-1.37}^{2.17}$ & $71.63_{-14.70}^{18.62}$ & $150.42_{-2.59}^{5.00}$ & $14.44_{-3.44}^{1.66}$ & $60.31_{-5.13}^{5.20}$ \\
FDR & 12 & $129.33_{-1.24}^{1.31}$ & $4.20_{-1.37}^{2.17}$ & $71.63_{-14.70}^{18.62}$ & $150.42_{-2.59}^{5.00}$ & $14.44_{-3.44}^{1.66}$ & $60.31_{-5.13}^{5.20}$ \\
\hline
\end{tabular}
%\caption{The distribution of fit parameters for Figure 1}
\label{Table_binning_rule}
\end{table*}

%%%%================================
\begin{figure*}
\centering
\includegraphics[width=0.7\linewidth]{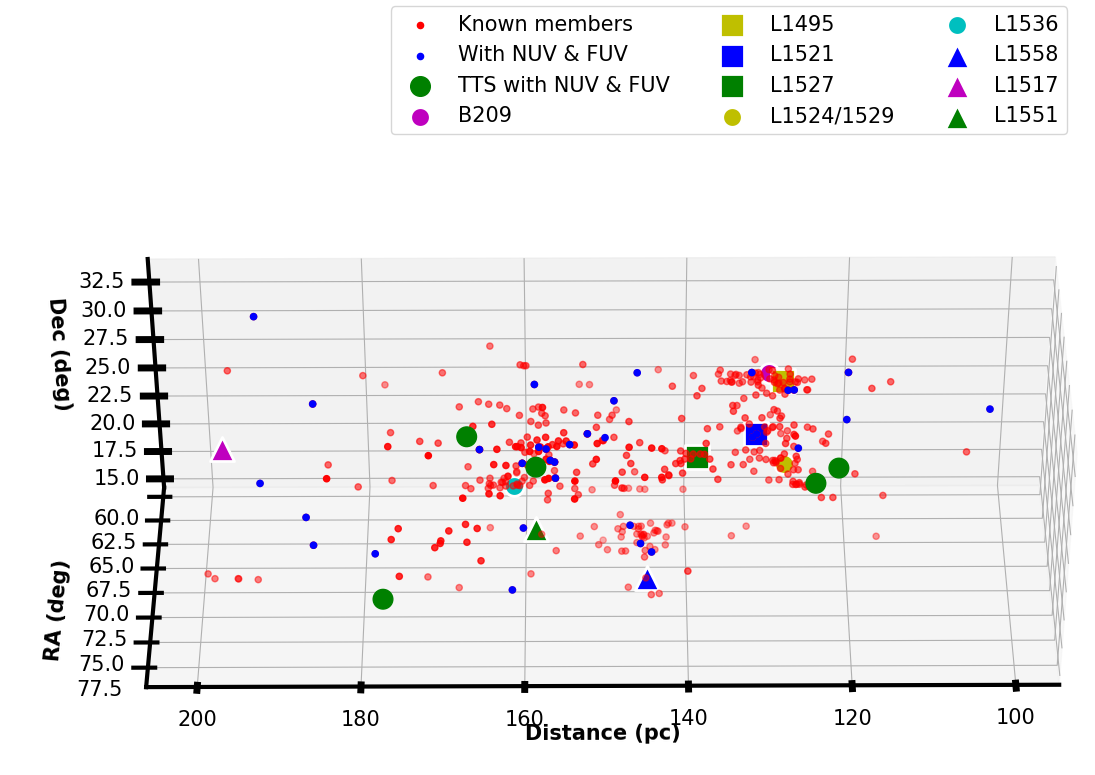}
\caption{ 3D representation of clouds and stars in the TMC. The names of the known clouds are represented in the plot. The distribution of newly added TTS are also shown. }
\label{Fig3}
\end{figure*}
%%%%%%====================================

\subsection{Application of binning methods to the sample of TMC candidates} \label{bimodality}

To convert the parallax to distance, we used the distance measurements estimated from \cite{BJ2018}. Also, we considered stars that have a distance range of 100 to 200 pc for this study.   
Using the various binning techniques discussed in the previous section, we analyze two samples of stars. This is shown in \autoref{Fig2}. The bimodal distribution in the distance for the young stars belonging to TMC is evident from \autoref{Fig2}. A similar distinct grouping of stars was reported by \citep{Fleming2019}. They, however, constructed the histograms by grouping the sources in 3 pc bins without explaining the justification for the bin size and the number of bins. They further used the sample from \cite{Rebull11}, which only has 159 stars, almost half the number of \cite{Luhman2018}. From \autoref{Fig2}, it is evident that the width and peak values change with various binning schemes. Hence, a proper analysis of the sample with various binning schemes is presented in the following section. 

In order to identify the peaks and width of the bimodal distribution, we fitted the histogram distribution with double Gaussian. We used the Markov chain Monte Carlo method implemented using the  $emcee$ python package \citep{EMCMC} to fit the double Gaussian. We fitted a double Gaussian of the form :

\begin{equation}
    f(x)=A_1 \times e^{\frac{-(x-\mu_1)^2}{2 \times \sigma_1^2 }} + A_2 \times e^{\frac{-(x-\mu_2)^2}{2 \times \sigma_2^2 }}
\end{equation}

where x is the distance in parsec, $A_1$ and $A_2$ are the peak amplitudes, $\mu_1$ and $\mu_2$ are the means of the two Gaussians and $\sigma_1$ and $\sigma_2$ are the standard deviations of the two Gaussians. We assumed a prior distribution of $A_1$ and $A_2$ to be uniform between 0--150, for  $\mu_1$ and $\mu_2$, we had a uniform distribution between 100--200 pc, while for the standard deviations  $\sigma_1$ and $\sigma_2$  we assumed a uniform distribution between 0--40 pc. The result of this analysis is tabulated in \autoref{Table_binning_rule}. It can be seen from \autoref{Fig2} and \autoref{Table_binning_rule} that the nearby population of stars (distance $<$ 140 pc) shows a broader distribution with higher amplitude, whereas the far population show less $\sigma$ with lower amplitude. 

To investigate how this bimodality in distance distribution translates to the structure of TMC, we analyzed the 3-dimensional distribution of stars in TMC from the \cite{Luhman2018} sample. This is shown in \autoref{Fig3}, also shown are the prominent dark clouds identified in TMC, namely B209, L1495, L1517, L1521, L1524, L1527, L1536, L1551, and L1558 (with the position and distance compiled from \citep{Luhman2018}). From the 3D representation, it is evident that the structure of the cloud is not strictly bimodal, as seen from the distance distribution. We see that there is a large clumpy of stars at a distance of 120-140 pc and a slightly diffused clumping of stars around 160-180 pc (\autoref{Fig3}). 

The near population of stars is found to align with the prominent clouds B209, L1495, and L1524. However, the stars contributing to the far peak in distance distribution are spread out over wide scales. The larger width values seen in the population of stars at the far end of the cloud is due to the fact that they are spread over a large distance. Also, this population of stars is sparsely populated in each of the clouds. In the case of near population, more stars are present, and they are concentrated spatially.

\begin{figure}
\centering
\includegraphics[width=1\linewidth]{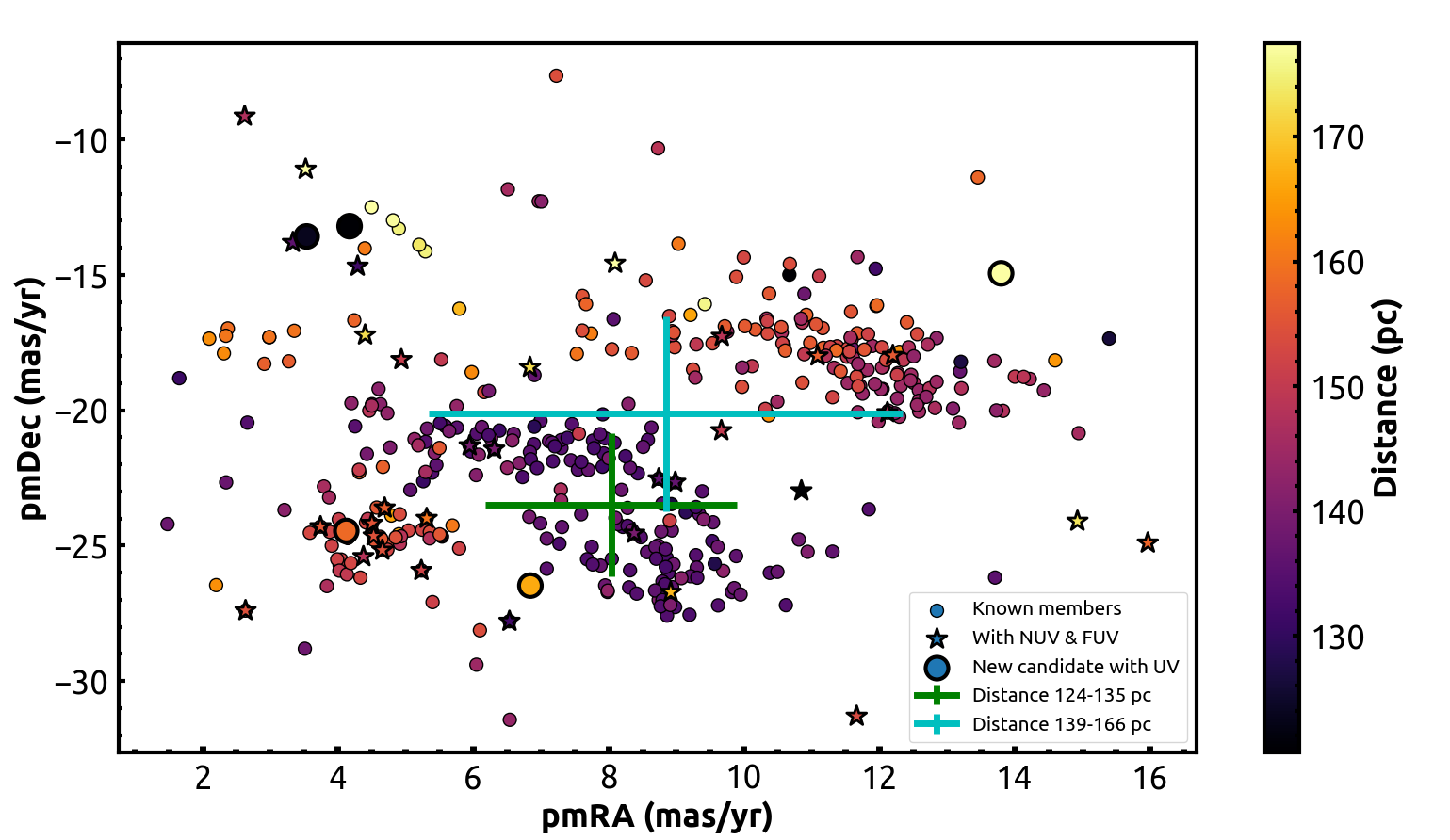}
\caption{ Proper motion diagram for the TMC members in \citep{Esplin2019} color coded with distance. A clear distinction in proper motion value with distance cannot be seen in the figure. The average values of proper motion for stars belonging to the two peaks are shown as green (distance between 124-135 pc) and cyan (distance between 139-166 pc) crosses. The standard error in the mean is used to represent the dispersion in proper motion for stars belonging to both peaks. The proper motion distribution of new TTS members are also highlighted with bigger filled circle.}
\label{Fig4}
\end{figure}

\begin{figure}
\centering
\includegraphics[width=0.55\textwidth]{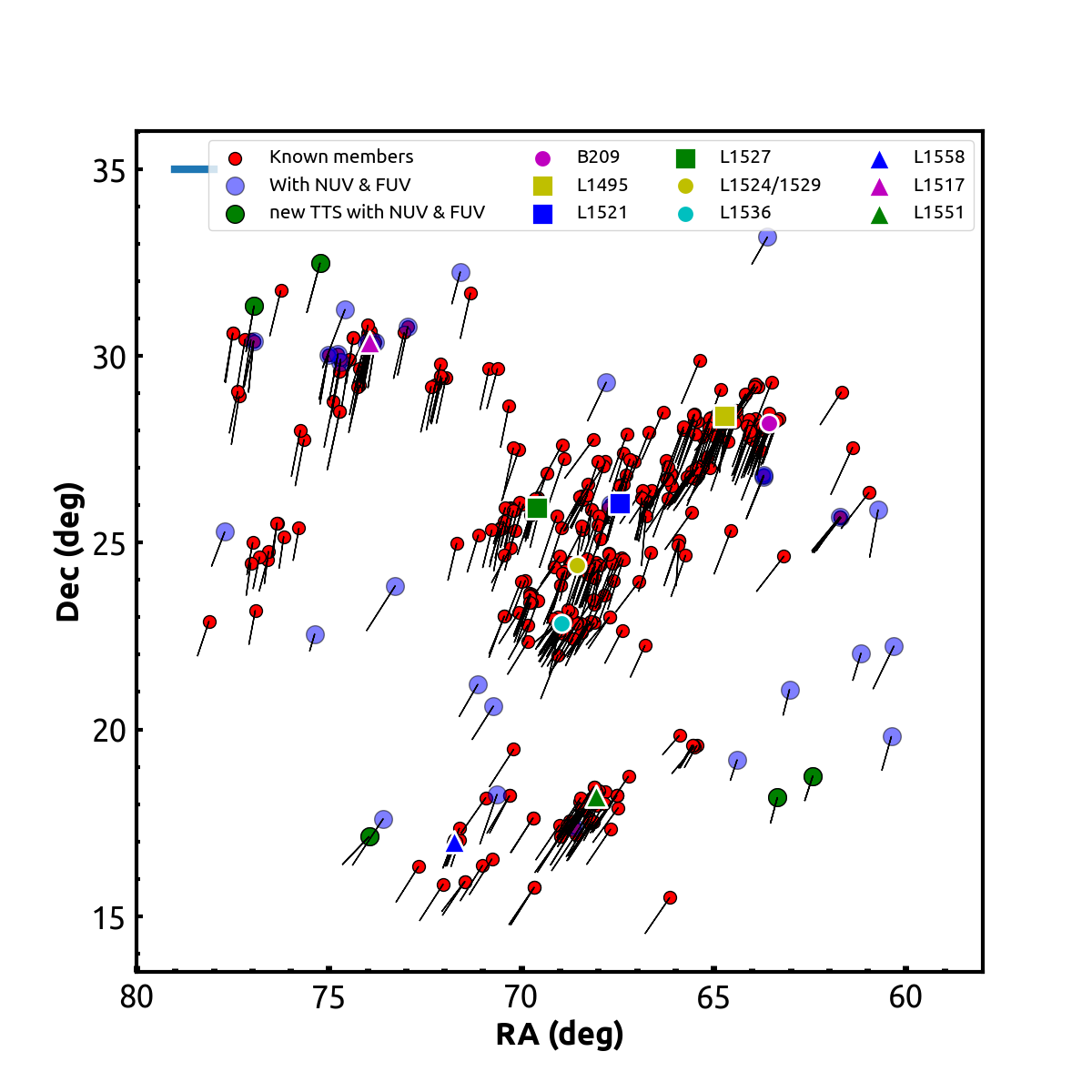}
\caption{ The position as well as proper motion of known members of TMC along the new TTS candidates in this study.  We have also shown the sources that have both NUV and FUV also detected. The blue horizontal bar on the top left represents the proper motion of 20 mas/yr.  The major Lynd clouds and Bernard clouds are also marked. }
\label{vector_point}
\end{figure}

\subsection{Proper Motion analysis} \label{proper_motion}

{ From Gaia DR3, we have accurate proper motion values of 310 stars in TMC. The distribution in proper motion in RA and DEC for the sample of stars is shown in \autoref{Fig4}. Based on the discussion in section 3.2, we divide the sample into two distance bins. From \autoref{Fig4}, one can also see that the spread in the distribution of proper motion of near and far members in the TMC is different. The near members show less spread in the proper motion values when compared to those at the far end of TMC. In order to quantify the distribution, we have included $16^{th}$ and $84^{th}$ percentile as a representation for dispersion in proper motion. This means that stars at the far end of the TMC distribution show a large range of velocity values, which is not seen with the near population. 
This suggests that the morphological distinction seen in the distribution of stars in the near and far end of the TMC (seen in \autoref{Fig2} and \autoref{Fig4}) is related to the distinction in dynamics. 
However, the mean proper motions of both the near and far ends of the TMC are similar (see \autoref{Fig4}). 
This stands in contrast with the findings of \cite{Fleming2019}, where they reported that the two populations had two distinct proper motion values. However, as stated before, the analysis done in \cite{Fleming2019} was for a much smaller sample of Taurus members. With the improved proper motion values from Gaia DR3 and a much larger sample, we do not see two distinct populations on the bases of kinematics. In \autoref{vector_point}, we have shown the proper motion vector diagram of the Taurus members, which also suggests that sources are moving in the same direction. 

In \autoref{Fig4}, we have also highlighted the distribution of new TTS and sources with UV detection. We notice that the proper motion of new TTS has a spread similar to the spread found in the far-end population. We can see that most of the sources with UV detection also follow the same spread. This is an artifact due to the lack of GALEX data in most of the Taurus region, mainly avoiding the region with a dense population due to the instrument's safety. This aspect gets more clear in the proper motion vector diagram in \autoref{vector_point}, where we can see the spatial distribution of the UV detected sources is mostly spread all over but not in the clumpy region. However, the figure shows that the kinematics of the UV-detected sources including the new TTS are similar to the rest of the populations. }

\begin{figure}[h]
\centering
\includegraphics[width=1\linewidth]{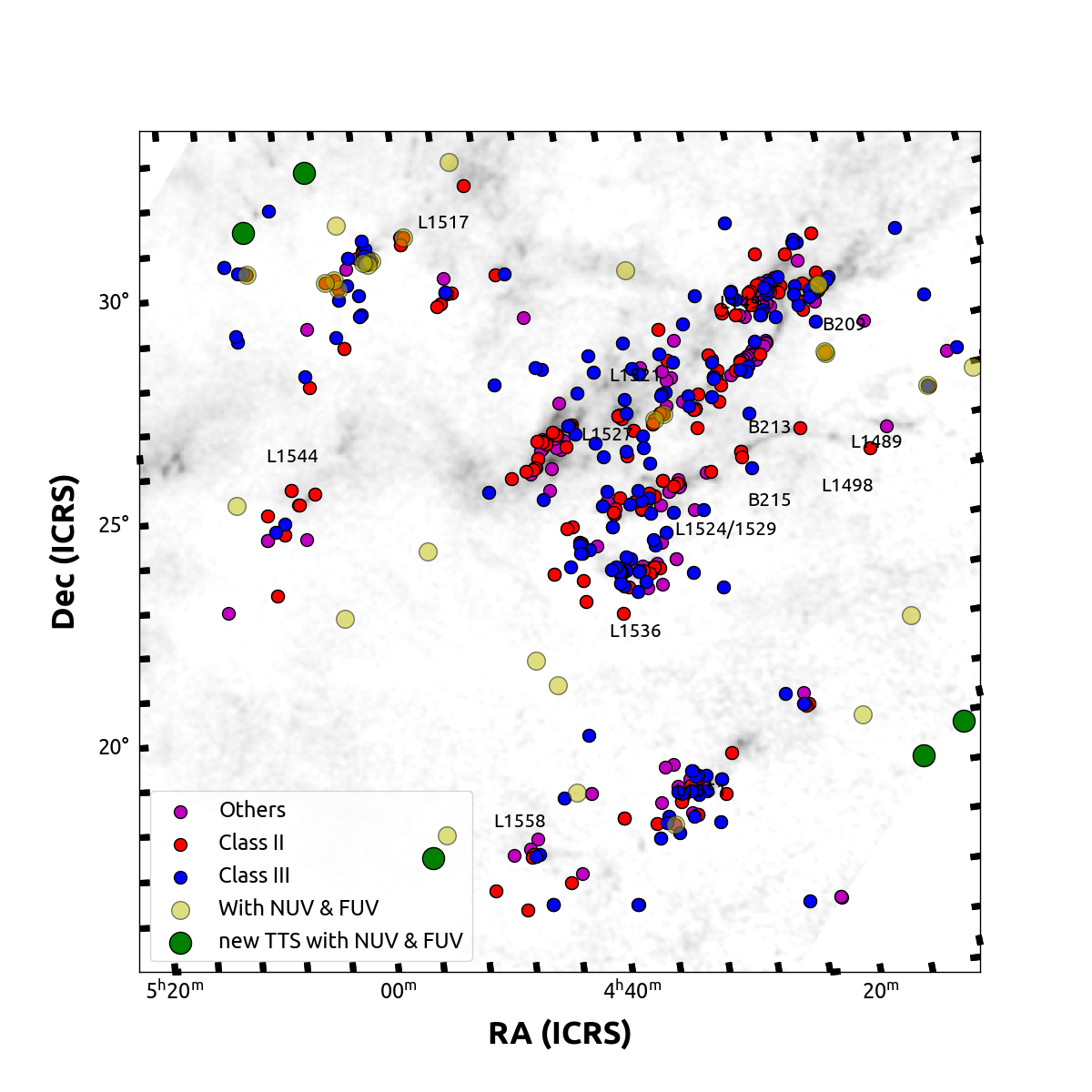}
\caption{ The spatial distribution of the known members of TMC (separated into different evolution stages) along the new TTS candidates in this study overplotted on the extinction map of TMC from \cite{Dobashi05}.  We have also shown the sources that have both NUV and FUV also detected. The major Lynd clouds and Bernard clouds are also marked. 
} 
\label{Fig5}
\end{figure}

\subsection{Spatial and proper motion analysis of Class II and Class III sources} \label{classII_III}

In the above analysis, we have used Class II and Class III sources. We, however, further wanted to investigate whether the distinction in the distance and proper motion for the sources in TMC is connected to its evolutionary phase. It has been suggested that Classical T Tauri stars (CTTS) and weak-line T Tauri stars (WTTS) are found over distinct distance ranges in TMC, with the former found between 126 and 173 pc, whereas WTTS is found on both sides of the molecular clouds, between 106 and 259 pc \citep{BG6}. This suggests that Class II and Class III sources are distributed in different manners. To assess this, we used the sample of 519 stars from \cite{Esplin2019}. The spatial distribution of the sample is shown in \autoref{Fig5}. We do not see any difference in the spatial distribution of Class II and Class III populations. The sources with UV detection are found to be spread mainly in the outskirts, avoiding the central dense star forming clouds and dense clumps, which is due the lack of GALEX observations in those regions. We also notice that many UV detected sources are mostly overlap with Class II sources, indicating that they might be actively accreting material from their circumstellar disks. With the current UV telescope UVIT and upcoming INSIST, we can get more of UV coverage of the Taurus region and details analysis of different class of populations and TTS.  

\begin{figure}
\centering
\includegraphics[width=1.0\linewidth]{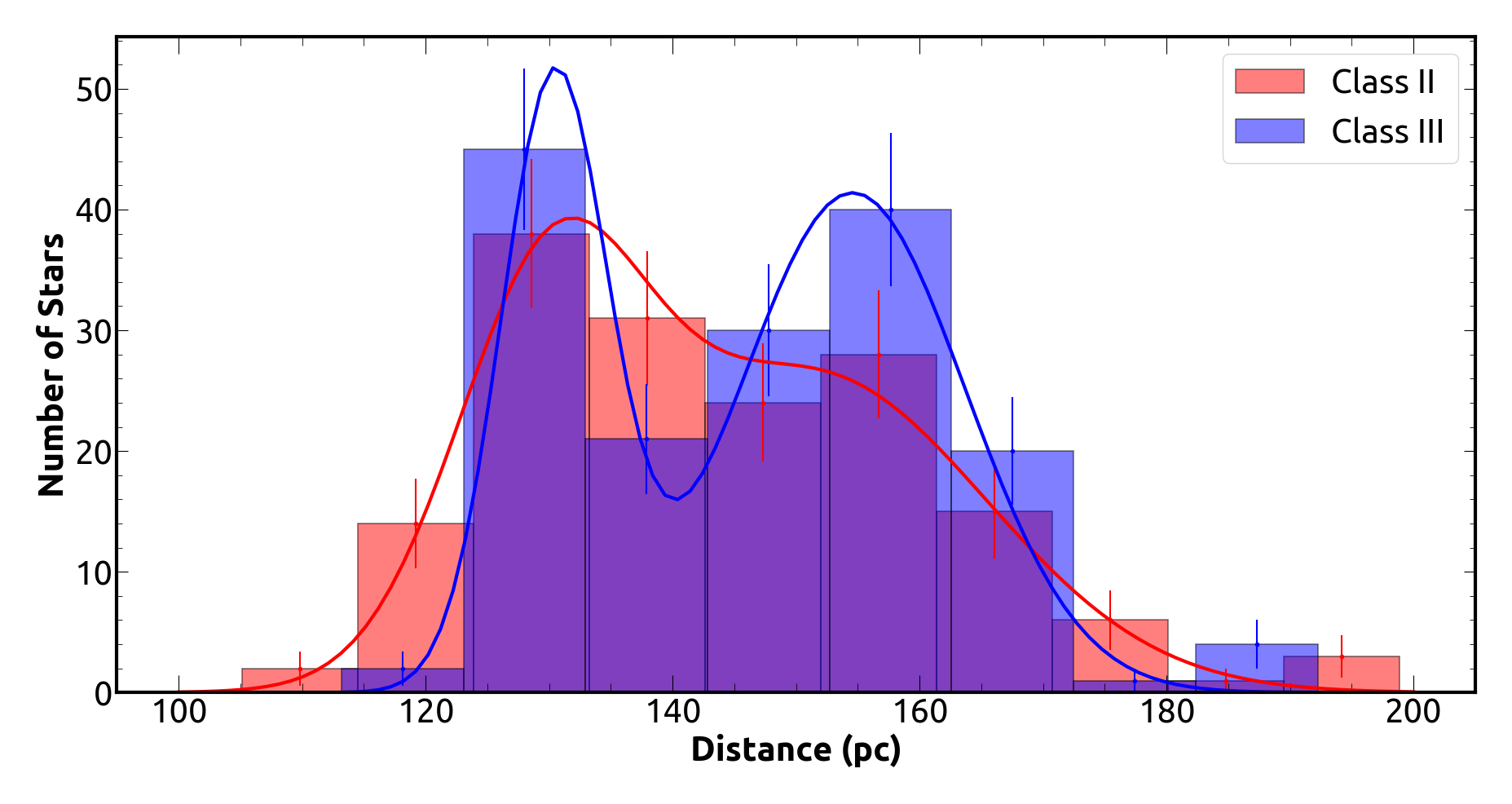}
\includegraphics[width=1.05\linewidth]{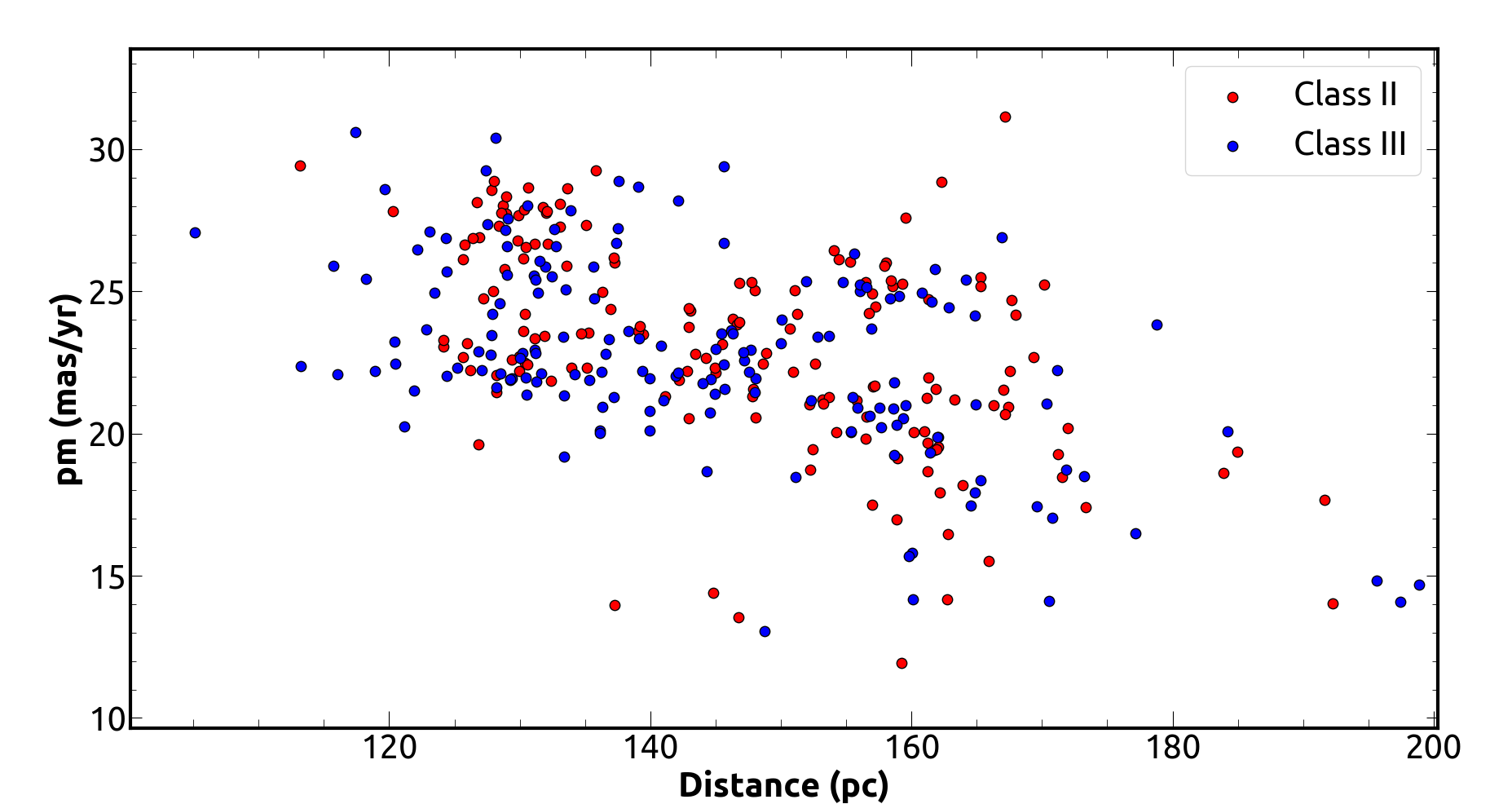}
\caption{(a) The distribution to the distance of the Class II (red) and Class III (blue) sources from the \cite{Esplin2019} sample is shown. Also shown is the fit to the distribution. (b) The proper motion of the Class II  (red points) and Class III (blue points) sample from \cite{Esplin2019} as a function of the distance.}
\label{Fig6}
\end{figure}

We then analyzed the distance distribution of Class II and Class III sources are shown in \autoref{Fig6}a. It is evident from \autoref{Fig6}a that both Class II and Class III sources independently show two peaks in the distance distribution and the it is more prominent for Class III. This agrees with \autoref{Fig2}, where we included the whole sample of stars for the analysis. We find that Class II sources show distance distribution peaking at $131.23_{-1.86}^{1.87}$ pc and $157.9_{-2.87}^{2.59}$ pc, whereas Class III sources peak at $127.98_{-1.40}^{1.82}$ pc and $158.7_{-4.38}^{3.48}$ pc. The 2-sample KS test shows that the probability of drawing the two distance distributions from the same sample is 52 $\%$.  

The proper motion for the Class II and Class III sources are calculated from $\mu_\alpha$ and $\mu_\delta$ values. The proper motion values are plotted with distance in \autoref{Fig6}b. The proper motion values of the sources span a range from 10 mas~yr$^{-1}$ to 35 mas~yr$^{-1}$. However, most of the sources are found to be around 22 mas~yr$^{-1}$, agreeing with previous studies \citep{BG6}. It can be seen from \autoref{Fig6}b that Class II and III sources are indistinguishable in terms of proper motion values. The two-dimensional two-sided KS test reveals that the associated probability that the two samples are drawn from the same populations is 17$\%$. Hence, we conclude that the distribution of Class II and Class III sources in TMC are similar both spatially and kinematically.

%%%==================================
\subsection{{\it Gaia} Color-Magnitude Diagram} \label{gaia_cmd}

\begin{figure}
\centering
\includegraphics[width=1\linewidth]{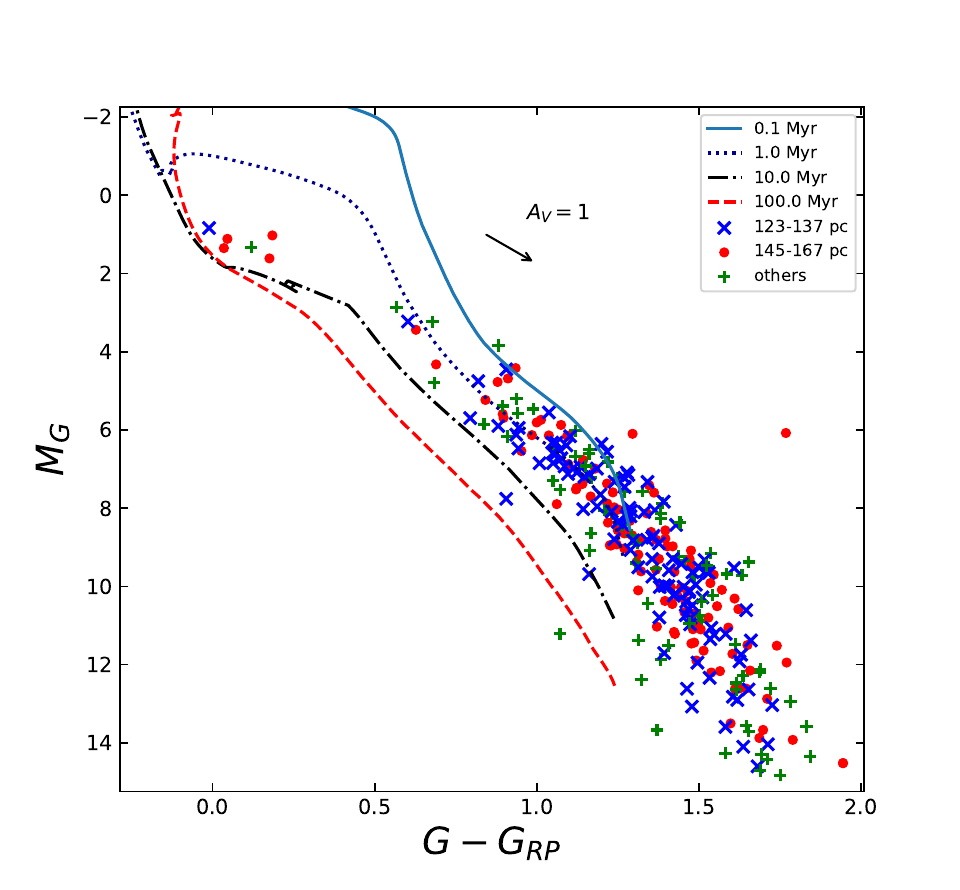}
\caption{Gaia CMD containing 304 TMC members over-plotted with the MIST isochrones of ages 0.1, 1, 10 \& 100 Myr. The blue crosses are the stars in the distance range 123-133 pc, and the red circles are stars at 140--161 pc. The green plus symbols are the stars that do not belong to near or far populations in the distance histogram. We used the MIST isochrones with metallicity, Z = 0.0152 and (V/V\textsubscript{crit}) = 0.4.}
\label{Fig7}
\end{figure}

In order to check whether the two populations seen in the distance (\autoref{Fig2}) translate to distinction in stellar parameters, we plotted the sample of stars in {\it Gaia} CMD. We found {\it Gaia} magnitudes for 304 stars from \cite{Luhman2018}, in the distance range 100--200 pc. The sample of stars belonging to two sub-populations are denoted in distinct colors. The sample of the near population belongs to the distance range 123--137 pc, whereas the far population is from 146--167 pc (considering the FDR scheme). The stars which do not belong to both these populations are mentioned as `others' in \autoref{Fig7}. We have not corrected the CMD for extinction since the reddening vector and the sequence of stars in TMC CMD are parallel \citep[see][]{Esplin2019}. The reddening vector is represented in \autoref{Fig7}. 

Over plotted in this figure are the isochrones for age estimation. `Modules for Experiments in Stellar Astrophysics (MESA) isochrones and evolutionary tracks' (MIST, \citealt{C16}) are used for the present analysis. MIST isochrones of ages 0.1, 1, 10, and 100 Myr are plotted along with the main sequence from \cite{PM13}. It can be seen that the whole sample of stars is found to be in the age range of 0.1--10 Myr. We are not able to fit the fainter and the redder part of the CMD as these the probably Brown Dwarf, and MESA does not reach that lower mass limit. Interestingly, there is no distinction between the location of stars belonging to the near and far ends of TMC.

%%%%================================
%%%%%==============================

\section{Conclusion}
 
 In this work, we have developed a new method using UV and optical CMD to identify TTS candidates. A large photometric survey of star-forming regions with UV telescopes (like UVIT and INSIST) combined with Gaia data will reveal many more new TTS candidates. % using our method. 
{  We do see two peaks in the distance distribution of Taurus members, but they are not very distinct, as suggested by \cite{Fleming2019} and named as two ´horns'. We notice that the far end has a relatively larger spread over distance compared to the near-end population. \cite{Fleming2019} reported that the mean proper motion of these two 'horns' was also different, where the authors used Gaia DR2 data. Using Gaia DR3 with a significant improvement in astrometry over Gaia DR2, we find that the mean proper motion of both the far and near-end population are similar and the far-end population has a relatively large spread in proper motion compared to the near-end.  
 We conclude that we do not find a distinction in the kinematics of stars at the near and far end of the TMC. Since the location of these young stars is correlated with the location of the dark (Lynd \& Barnard) 
 clouds, such a distinction is connected to the star formation process in TMC. We also found that both populations have similar age distributions.}

 The major results of this study are summarized below. 
\begin{itemize}

    \item We have developed new criteria to identify new TTS candidates using UV and optical CMDs by combining the GALEX and Gaia surveys. We found 5 new Taurus members, which are newly identified TTS candidates in the TMC. 
    
    \item The distance distribution of {\it Gaia} sources in the TMC is found to have two peaks, at $130.17_{-1.24}^{1.31}$ and $156.25_{-5.00}^{1.86}$ pc (from the FDR method), respectively, but not very distinct as reported previously. This may suggest a dumbell-type of morphology in the distance distribution of sources in TMC. However, from the three-dimensional representation of the sources in RA-DEC-distance space, we found that the sources at the far end of the TMC ($>$140 pc) are more scattered when compared to the nearby sources. 
    
    \item From the proper motion analysis, we do not find a distinction in dynamics between the near and far sources in the TMC, however, the sources on the far end have a larger spread in proper motion compared to that found on the near end.
    %matching the morphological scenario.  
    
    \item The spatial and kinematic distribution of Class II and Class III sources is very similar. 
    
    \item The distinction in the spatial location of stars does not translate to a difference in age, as seen from their distribution in the {\it Gaia} color-magnitude diagram.

\end{itemize} 

\section{Acknowledgment}

We thank all the people that have made this possible. We thank the staff of DOT, IAO, Hanle, and CREST, Hosakote, that made spectroscopic observations possible. The facilities at DOT are operated by ARIES, Nainital. The facilities at IAO and CREST are operated by the Indian Institute of Astrophysics, Bangalore.
PKN acknowledges TIFR’s postdoctoral fellowship. PKN also acknowledges support from the Centro de Astrofisica y Tecnologias Afines (CATA) fellowship via grant Agencia Nacional de Investigacion y Desarrollo (ANID), BASAL FB210003. 
This research has also made use of NASA’s Astrophysics Data System Abstract Service and of the SIMBAD database, operated at CDS, Strasbourg, France. This work presents results from the European Space Agency (ESA) space mission, Gaia. Gaia data are being processed by the Gaia Data Processing and Analysis Consortium (DPAC). Funding for the DPAC is provided by national institutions, in particular, the institutions participating in the Gaia MultiLateral Agreement (MLA). The Gaia mission website is https://www.cosmos.esa.int/gaia. The Gaia archive website is https://archives.esac.esa.int/gaia.

% \usepackage{longtable}
% Note: It may be necessary to compile the document several times to get a multi-page table to line up properly

% \bibliographystyle{apj}

%\usepackage{appendix}

\appendix \label{appendix}

%As mentioned in the section 3

\section{Sturges' Rule}

Sturges' rule \citep{Sturges26} provides a simple method of binning the data based on the number of data points, $n$. Sturges' rule makes the assumption that the histogram consists of normally distributed data points, approximated as a binomial distribution. The optical number of bins, $k$, using the Sturges' rule \citep{Sturges26} is given as 

\begin{equation}
    k= 1 + \log_2( n )
\end{equation}

\section{Rice Rule and the rule of Square roots}

Other methods based on the number of data points in the sample are the Rice rule \footnote{http://onlinestatbook.com/Online\_Statistics\_Education.pdf} and the `rule of square roots. The number of bins, $k$. as given by `Rice rule' is $k=\sqrt[3]{n}\times2$ and from the `rule of square roots' is $k=\sqrt{n}$. The issue with these methods is that they do not consider the inherent skewness of the data. 

\section{Doane's rule}

 If the data points are not normally distributed, we can use a modification of the Sturges' rule, known as Doane's rule \citep{Donae76}. Doane's rule tries to account for the skewness of the data. The number of bins, $k$, is given as :

\begin{equation}
     k = 1 + \log_2( n ) + \log_2 \left( 1 +  \frac { |g_1| }{\sigma_{g_1}} \right) 
\end{equation}

 where  $g_{1}$ is the estimated 3rd-moment-skewness of the distribution and is  given as

\begin{equation}
    g_{1}=\frac{\Sigma_{i=1}^{n} (X_i - \bar{X})^3}{[\Sigma_{i=1}^{n} (X_i - \bar{X})^2]^{3/2}}
\end{equation}

Here $X_i$ are the data points, and $\bar{X}$ is the mean of the sample. $\sigma _{g_{1}}$ is given as:

\begin{equation}
    { \sigma _{g_{1}}={\sqrt {\frac {6(n-2)}{(n+1)(n+3)}}}} 
\end{equation}

\section{Scott's rule}

`Scott's rule' \citep{Scott79} is based on the standard deviation, $\sigma$, of the data. Unlike previous methods,  Scott's rule gives us the optimal bin width, $h$, and not the number of bins, $k$. The bin width, $w$, is correlated with the number of bins by $k=\frac{R}{w}$, where $R$ is the range in the distribution. The bin width is given as follows:  

\begin{equation}
    h = \frac{3.49  \sigma}{\sqrt[{3}]{n}}
\end{equation}

\section{Freedman-Diaconis rule}

The Freedman-Diaconis rule \citep[hereafter FDR;][]{FDR} has a similar approach as that of Scott's rule, wherein the optimal bin width is considered instead of the number of bins. The optimal number of bins in the FDR rule depends on the interquartile range ${\text{IQR}}(x)$. The ${\text{IQR}}(x)$ is less sensitive to deviant outlier points than the usual standard deviation estimations. Standard deviation calculation depends on the mean of the data where outlier data points are included. This may affect the accuracy of the result, and hence ${\text{IQR}}(x)$ will be the best way to estimate the bin size. The optical bin width is estimated as  

\begin{equation}
    h=2\,{{\text{IQR}}(x) \over {\sqrt[{3}]{n}}}
\end{equation}

\end{document}